%% file: main.tex
\documentclass[sigconf,authorversion,nonacm]{acmart}
\usepackage{multirow}
\usepackage[title]{appendix}
\usepackage{rotating}
\usepackage[linesnumbered,,ruled]{algorithm2e}
\usepackage[textsize=tiny, textwidth=2cm]{todonotes}
\usepackage{url}
\usepackage{diagbox}
\usepackage{amsmath}
\usepackage{graphicx}
\usepackage{subfig}
\usepackage{colortbl}
\usepackage{enumitem}

\AtBeginDocument{%
  \providecommand\BibTeX{{%
    \normalfont B\kern-0.5em{\scshape i\kern-0.25em b}\kern-0.8em\TeX}}}
    
\begin{document}

\title{DMSConfig: Automated Configuration Tuning for Distributed IoT Message Systems Using Deep Reinforcement Learning}

\author{Zhuangwei Kang}
\authornotemark[1]
\affiliation{%
  \institution{Vanderbilt University}
  \city{Nashville}
  \state{Tennessee}
  \country{USA}
}
\email{zhuangwei.kang@vanderbilt.edu}

\author{Yogesh D. Barve}
\affiliation{%
  \institution{Vanderbilt University}
  \city{Nashville}
  \state{Tennessee}
  \country{USA}
}
\email{yogesh.d.barve@vanderbilt.edu}

\author{Shunxing Bao}
\affiliation{%
  \institution{Vanderbilt University}
  \city{Nashville}
  \state{Tennessee}
  \country{USA}
}
\email{shunxing.bao@vanderbilt.edu}

\author{Abhishek Dubey}
\affiliation{%
  \institution{Vanderbilt University}
  \city{Nashville}
  \state{Tennessee}
  \country{USA}
}
\email{abhishek.dubey@vanderbilt.edu}

\author{Aniruddha Gokhale}
\affiliation{%
  \institution{Vanderbilt University}
  \city{Nashville}
  \state{Tennessee}
  \country{USA}
}
\email{a.gokhale@vanderbilt.edu}

\renewcommand{\shortauthors}{Kang, et al.}

\input{abstract}
\input{ccs-concepts}


\maketitle

\input{introduction}
\input{problem}

\input{dmsconfig}

\input{experiment}
\input{related}
\input{conclusion}
\input{acknowledgements}

\bibliographystyle{ACM-Reference-Format}
\bibliography{reference}

\appendix

\end{document}

%% file: abstract.tex
\begin{abstract}
Distributed messaging systems (DMSs), which are critical for the timely and reliable data dissemination needs of IoT systems, are typically equipped with a large number of configurable parameters that enable users to define application run-time behaviors and resource distribution rules. However, the resulting high-dimensional configuration space makes it difficult for users to determine the best configuration options that can maximize application throughput given specific latency constraints, which is referred to as the constrained configuration tuning problem. Although many studies have been proposed to fulfill automatic software profiling, they have several limitations. First, they optimize throughput only, which assumes throughput and latency are strictly inversely proportional but this is not true for concurrent systems. Second, although some studies regard the optimization objective as a proportional summation of latency and throughput, the explicit latency limitation may not be consistently guaranteed. Third, existing approaches iteratively search the optimal configuration by discretizing parameter ranges, which may result in local optimum, however, most DMS parameters are continuous.

To overcome these challenges, we propose a novel, automatic configuration tuning system called DMSConfig, which combines conventional machine learning methods and popular deep reinforcement learning. DMSConfig explores the configuration space by interacting with a data-driven environment prediction model (a DMS simulator), which eliminates the prohibitive cost of conducting online interactions with the production environment. We develop an emulation-based testbed to accelerate the data collection process when building the DMS simulator. DMSConfig employs the deep deterministic policy gradient (DDPG) method and a custom reward mechanism to learn and make configuration decisions based on predicted DMS states and performance. Since the environment model is decoupled with our optimization objectives, DMSConfig is highly adaptive to serve tuning requests with different latency boundaries. We validated the performance of DMSConfig on a representative DMS (Kafka for this paper) under 9 use cases and 7 latency constraint conditions. Experimental results show that DMSConfig performs significantly better than the default configuration and has better adaptability to CPU and bandwidth-limited environment. We also achieve almost the same throughput as three prevalent parameter tuning tools, but with fewer violations of latency constraints. 

\end{abstract}

%% file: ccs-concepts.tex
\ccsdesc[500]{Software and its engineering~Software configuration management and version control systems}
\ccsdesc[500]{Computing methodologies~Policy iteration}

%% file: introduction.tex
\section{Introduction}
\label{sec:intro}

With the advancement of bandwidth infrastructures, it has become a trend to migrate Internet of Things(IoT) devices to fast networks (5G)\cite{kim2018road,minoli2019practical}, which is conducive to resolving performance degradation arisen by network instability and bandwidth limitations in IoT environments. Nevertheless, to leverage and augment the practical utility of 5G networks, it is vital to ensure the rationality of the design and configurations of IoT software at the application level first. Many IoT application domains, such as smart cities and smart grids, employ distributed message system(DMS) as the middleware for data transmission through which data samples can be produced, disseminated and consumed asynchronously. To ensure flexibility in a wide range of deployment scenarios, system topologies and runtime specifications, industrial-strength DMS provide users with a set of configurable parameters that have different data types (e.g., numeric, boolean, categorical) and value ranges, which consequently constitute a hybrid, multidimensional configuration space. These parameters control application runtime behaviors and resource allocation strategies resulting in different variations of application performance measured across different metrics, such as throughput, latency, CPU utilization, etc. Although DMS software is typically equipped with a default profile for the sake of general applicability, the default profile may not consistently guarantee optimal performance under a wide range of deployment use cases and QoS constraints(justify in section~\ref{sec:exp}). Making prudent configuration decisions requires fine-grained domain knowledge and in-depth understanding of the impact of each parameter on application performance, as well as their unseen interactions, which is challenging even for experts, not to mention common users. Besides, na\"ive exhaustive search methods are laborious, time-consuming, suboptimal and not scalable. Hence, this paper delves into developing an intelligent and automated configuration recommendation system for DMS that aims to optimize application throughput with specific latency constraints, which meets the demand of practical IoT streaming applications that usually have stringent requirements on both throughput and response time. We formally term the problem as a constrained configuration optimization problem(more details are available in Section~\ref{sec:problem}) and choose Apache Kafka~\cite{garg2013apache} as an example to illustrate our approach as it is one of the most popular DMSs in the last decade. Note that since our approach is not tied to Kafka and its system components are fully decoupled, it is applicable to other DMS implementations as well.


The published literature\cite{herodotou2020survey,huang2019survey,xu2015systems} suggest that there are three representative classes of approaches in the state-of-the-art configuration tuning domain: 

\textbf{Search-based Methods (SBM)}, as in \cite{zhu2017acts,wang2013searching,murashkin2013visualization}, iteratively evaluate candidate configurations derived from heuristic rules and historical experience, and gradually narrow the search space based on observed application performance. Such online search algorithms require less tuning time; however, they incur the following limitations: (1) designing a practical heuristic function is challenging, particularly if the objective function is subjected to some conditions; (2) the heuristic algorithms cannot guarantee global optimum; and (3) the entire search process needs to be restarted if the optimization objective or constraint condition is out of date. 

\textbf{Prediction-based Methods (PBM)}, like \cite{van2017automatic,sarkar2015cost,nair2017using,mahgoub2017rafiki}, are the most straightforward and general approaches that first build an offline performance prediction model based on a pre-collected training dataset and then couple the learned model with some heuristic searching strategies (e.g., genetic algorithms or recursive random search) to discover the near-optimal configuration. The pre-trained model acts as a simulator of the target application during the tuning process through which the search engine can rapidly explore the configuration space. Thus, PBM is more likely to discover the near-optimal configuration. Such data-driven methods also enable the tuner to quickly adapt different optimization objectives and QoS constraints with no need to re-interact with the target application. Despite the above advantages, the most critical barrier in adopting PBM is building an accurate performance prediction model because it needs a large volume of training data, which is hard to collect in production environments. 

\textbf{Rank-based Methods (RBM)}, such as \cite{bao2018autoconfig, zhu2017bestconfig, gu2019multi}, cope with the challenges of PBM by reducing information granularity required by the searching stage. Rather than predicting the exact value of performance metrics, RBM builds a machine learning model to estimate the rank order of configurations. Since any two observations can form a training sample, $ ( \begin{smallmatrix} n\\2 \end{smallmatrix} ) = \frac{n(n-1)}{2}$, the training set is significantly augmented. However, RBM is infeasible in constrained optimization scenarios because such gray-box methods do not capture exact values of performance metrics during the training phase. Also, prior efforts study the continuous configuration space in a discrete manner; nonetheless, deciding the length of sub-intervals for each parameter in every iteration is rather hard since parameters typically have different importance, value range, and nonlinear correlation. 


To tackle the constrained configuration optimization problem and make up for the gaps in the above-mentioned approaches, we develop a novel DMS configuration recommendation system called \emph{DMSConfig} that combines container-based emulation techniques, conventional machine learning, and the deep deterministic policy gradient (DDPG\cite{lillicrap2015continuous}) reinforcement learning(RL) algorithm. DMSConfig falls in the category of PBM and contains three main stages: data collection, DMS simulator training, and configuration tuning. However, DMSConfig significantly alleviates the limitations in PBM.  For instance, in the data collection phase, we leverage container and traffic control techniques to emulate practical DMS production environments in terms of CPU, memory, and bandwidth, which enhance adaptivity of our system in practice. Moreover, owing to the resource isolation fulfilled by container virtualization, we can maximize the resource utilization rate of physical nodes and evaluate multiple configurations in parallel to significantly alleviate the burden of data collection cost. We adopt the random forest (RF) algorithm to train our DMS simulator that takes ten of the most performance-relevant parameters as input and forecasts five Kafka internal state metrics, publisher-side throughput, and latency. DMSConfig uses DDPG to find the near-optimal configuration under the latency constraint for the following reasons: (1) the sequential decision-making(SDM) process in RL coincides with the essence of iterative parameter adjustment; (2) DDPG has been proven a robust approach for settling continuous control problems(continuous configuration in our context); (3) the reward function in RL guides the tuning process by applying revenue or penalty to the agent, which satisfies our demand for throughput and latency simultaneously; (4) driven by the model-based DMS simulator and RL reward mechanism, DMSConfig can rapidly adapt tuning requests that have different latency constraints.

We conducted extensive experiments under 5 dimensions of external settings, including message size, producer amount, producer CPU number, and bandwidth. Experimental results show that in 9 experimental use cases, DMSConfig outperforms the default configuration by 12\%-538\% and guarantees near-optimal throughput performance under scenarios with limited CPU or bandwidth resources. Compared to three state-of-the-arts automated parameter tuning frameworks, DMSConfig provides analogous throughput, but more reliable latency assurance on 7 levels of latency constraints.

In summary, this paper makes the following contributions:
\begin{enumerate}
    \item We develop an intelligent and automated configuration recommendation system, which can automatically find the configuration that maximizes the producer-side throughput and meets latency restriction in IoT scenarios despite the presence of a continuous high-dimensional configuration space;
    \item We design an effective reward function for DDPG that accelerates the speed of locating high-quality Kafka configurations for latency-aware producers;
    \item We create a performance prediction model that can successfully estimate Kafka internal state and performance metrics for a given workload and system topology; and
    \item We develop a full-stack, container-based Kafka emulation and monitoring system that is practical for conducting emulation-based Kafka benchmark tasks.
\end{enumerate}


The remainder of the paper is organized as follows: Section~\ref{sec:problem} formulates the constraint configuration problem for DMS; Section~\ref{sec:dmsconfig} depicts the architecture and workflow of our DMSConfig solution; in Section~\ref{sec:exp}, we present the design of experiments, and compare the performance of DMSConfig, Kafka default configuration and three baseline approaches under multiple IoT use cases; Comparison with related research appears in Section~\ref{sec:related}; and finally, we summarize this paper and lessons learned in Section~\ref{sec:conclusion}.


%% file: problem.tex
\section{Motivation and Problem Formulation}
\label{sec:problem}

\begin{figure}[H]
    \centering
    \includegraphics[width=0.9\columnwidth]{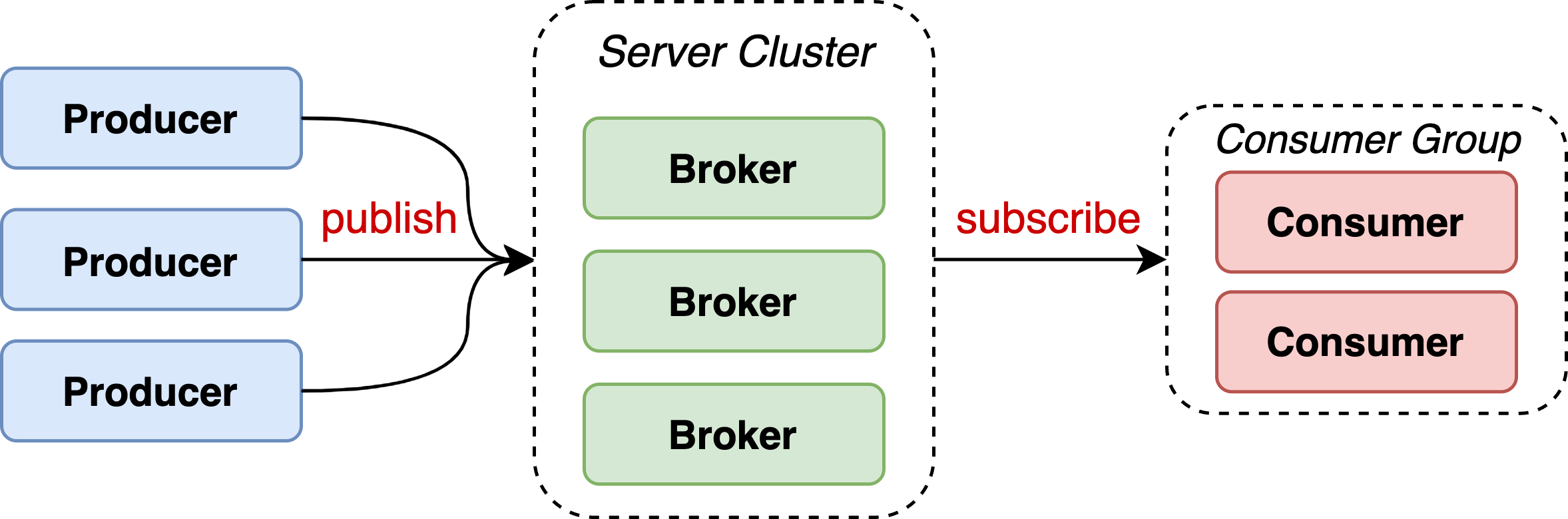}  
    \caption{\textmd{An example Pub/Sub Deployment}}
    \label{fig:pubsub}
\end{figure}

A distributed message system, as shown in Figure~\ref{fig:pubsub}, is composed of three  entity types namely: brokers, producers, and consumers. Brokers are responsible for receiving, persisting, and forwarding events dispatched by producers. The subscribing mechanisms on the consumer-side differ across DMS implementations, which may include pull-based (e.g., Kafka, RocketMQ, etc.) and push-based (e.g., RabbitMQ, ActiveMQ, etc.) subscription patterns. Therefore, to maximize system capacity and develop a general solution to automated configuration, we choose to optimize producer-side throughput, which is the amount of data that brokers can serve per unit time. On the other hand, although throughput and latency are inversely proportional in simple, synchronous applications, their relationship becomes fuzzy for concurrent systems with complex networking topologies as in IoT systems.  For instance, in DMS deployed in IoT networks, latency is jointly affected by many factors including system parallelism, message backlog, micro-batching, message retransmission, etc. Hence, there is a need to take bounded latency into account while optimizing producer throughput because the service-level objective is usually multifold for practical streaming workloads rather than only throughput. 

To this end, this paper delves into the constrained configuration problem for distributed message systems, shortened as the \emph{CCDMS} problem. Next, we define fundamental components of the CCDMS problem and formulate the problem formally.

\textbf{Configuration Space}: The fundamental unit of configuration space $CP$ is an individual parameter $p$, whose value is in a continuous real range (numeric parameters) or in a finite discrete set (boolean and category parameters). We designate boundaries of numeric parameters according to the Kafka manual\cite{ApacheKa30:online} and practical experience. Also, we index the value of a discrete parameter by its position in the array. Furthermore, we model one configuration as a vector $\vec{C}$ of parameters that are deployed when the application is launched and consider it immutable at runtime. 



\textbf{Resource Profile}: Resource profile $R$ involves information on the number of CPU cores $u$, available physical memory $m$, and free disk space $d$ assigned to each entity, and the overall bandwidth $b$ in the system. Collectively, we model it as a matrix $R=\{\vec{u}, \vec{m}, \vec{d}, \vec{b}\}$, where columns indicate resource types and rows reflect details of each entity.

\textbf{Workload and Topology}: Let $W=(f, l, s, r)$ be the unified workload snapshot for producers, where $f$ denotes the number of messages emitted by producers per second, $l$ is the length of the messages, $s$ indicates the sending mode of messages (synchronous or asynchronous), and $r$ refers to the transmission reliability (best effort or reliable). Further, we model application topology $T$ as $T=(N_p, N_b, N_c)$, referring to the number of producers ($N_p$), brokers ($N_b$), and consumers ($N_c$) in the application, respectively.

\textbf{Objective}: The goal of CCDMS is to maximize throughput $TP$ given a latency constraint $L_c$, where $TP$ denotes overall messages acknowledged by brokers per second, and $L_c$ is the restriction imposed on system latency (average latency of all producers). In summary, we formulate the CCDMS problem as follows:
\begin{equation}
\label{eq:obj}
\begin{aligned}
& \underset{C \in CP}{\text{max}}
& & TP(W, T, R, C) \\
& \text{s.t.}
& & latency \leq L_c
\end{aligned}
\end{equation}


Since most DMS parameters are continuous and searching for the optimal configuration in such a space is proven to be NP-hard \cite{sun2008complexity}, we cast the problem of CCDMS into an RL task that learns a trajectory approaching the optimal configuration through a trial-and-error strategy. RL is a branch of machine learning that is originally inspired by the behaviorist theory in psychology and has succeeded in many stochastic optimization problems. The theoretical basis for RL are Markov Decision Processes. Specifically, a Markov Decision Process is defined by a four-tuple $M=\{S, A, R, T\}$, where $S$ denotes all possible states in the environment and $A$ is the action space. The reward function $R \sim r(s_t, a_t)$ estimates the immediate revenue of taking an action $a_t\in A$ under state $s_t \in S$ at each decision epoch $t$. The transition of agent states is governed by a conditional probabilistic function $T=P\{s_{t+1}=s'|s_t=s, a_t=a\}$. There might be multiple executable actions with diverse transition probabilities for any state, yielding different subsequent states and long-term returns. By continuously interacting with the environment $\epsilon$, the agent aims to find an optimal policy $\pi^{*}$ by maximizing the expected accumulated discounted rewards over the future through taking effective actions in any state, denoted as $\pi^*: S\rightarrow A$. Therefore, the state value function under policy $\pi^*$ is:
\begin{equation}
\label{eq:val_func}
\begin{aligned}
& V_*(s) = \max_{\pi \in \Pi} V_{\pi}(s) \\
& = \max_{\pi \in \Pi} \ E_{\pi}\left[\sum_{t=0}^\infty \gamma^t r(s_t, a_t) | s_0=s\right], \forall s\in S
\end{aligned}
\end{equation}
, where $\gamma$ is the discount factor in range $\left[0, 1 \right]$, representing the trade-off between immediate and future rewards. A larger $\gamma$ indicates the agent is more interested in short-term revenues. Eq.~\ref{eq:val_func} implies that the state value at an arbitrary step is determined only by the instantaneous reward and the consecutive state value, thereby, the optimization criteria of MDP can be recursively represented as follows, known as Bellman Optimality Equation (BOE):
\begin{equation}
\label{eq:boe}
\begin{aligned}
& V_*(s) = \max_{a} \ E\left[r(s, a) + \gamma V_*(s')\right | s_0=s], \forall s\in S 
\end{aligned}
\end{equation}
The above equation shows that the original optimization problem is decomposed into seeking the best single-step action. The agent incrementally enhances the likelihood of selecting high-quality actions through trial and error until it can obtain the maximum state value in any state. Meanwhile, the exported optimal policy can navigate the agent by acquiring the maximum accumulated return in the terminal state. 

Accordingly, we need to map the CCDMS problem to MDP context for which we define key components as follows:
\begin{enumerate}
    \item \textbf{Environment:} The environment is the DMS being tuned. In our implementation, we use a pre-trained regression model to simulate a practice DMS environment, which takes a specific configuration as input and generates estimated DMS internal and external states. Note that the environment model is built based on a concrete workload, topology, and resource profile. 
    \item \textbf{Agent:} An intelligent configuration tuner based on a deep RL model is regarded as the agent.
    \item \textbf{Action:} Action is represented as a vector of tunable DMS parameters. 
    \item \textbf{State:} We use a group of observable DMS internal metrics, represented as a vector, to describe DMS states.
    \item \textbf{Reward:} Reward is the amount of throughput increment compared to that of the initial setting and the previous one. Further, we assign a penalty to the reward if the latency constraint is violated. More design details of the reward function are shown in Section~\ref{sec:dmsconfig}.
    \item \textbf{Policy:} In DMSConfig, we feed the agent state into a deep neural network, which then outputs a deterministic action that the agent should take. The policy network is iteratively updated based on the policy gradient method. 
\end{enumerate}

%% file: dmsconfig.tex
\section{DMSConfig Methodology}
\label{sec:dmsconfig}

\begin{figure}[htb]
    \centering
    \includegraphics[width=\columnwidth,height=8cm]{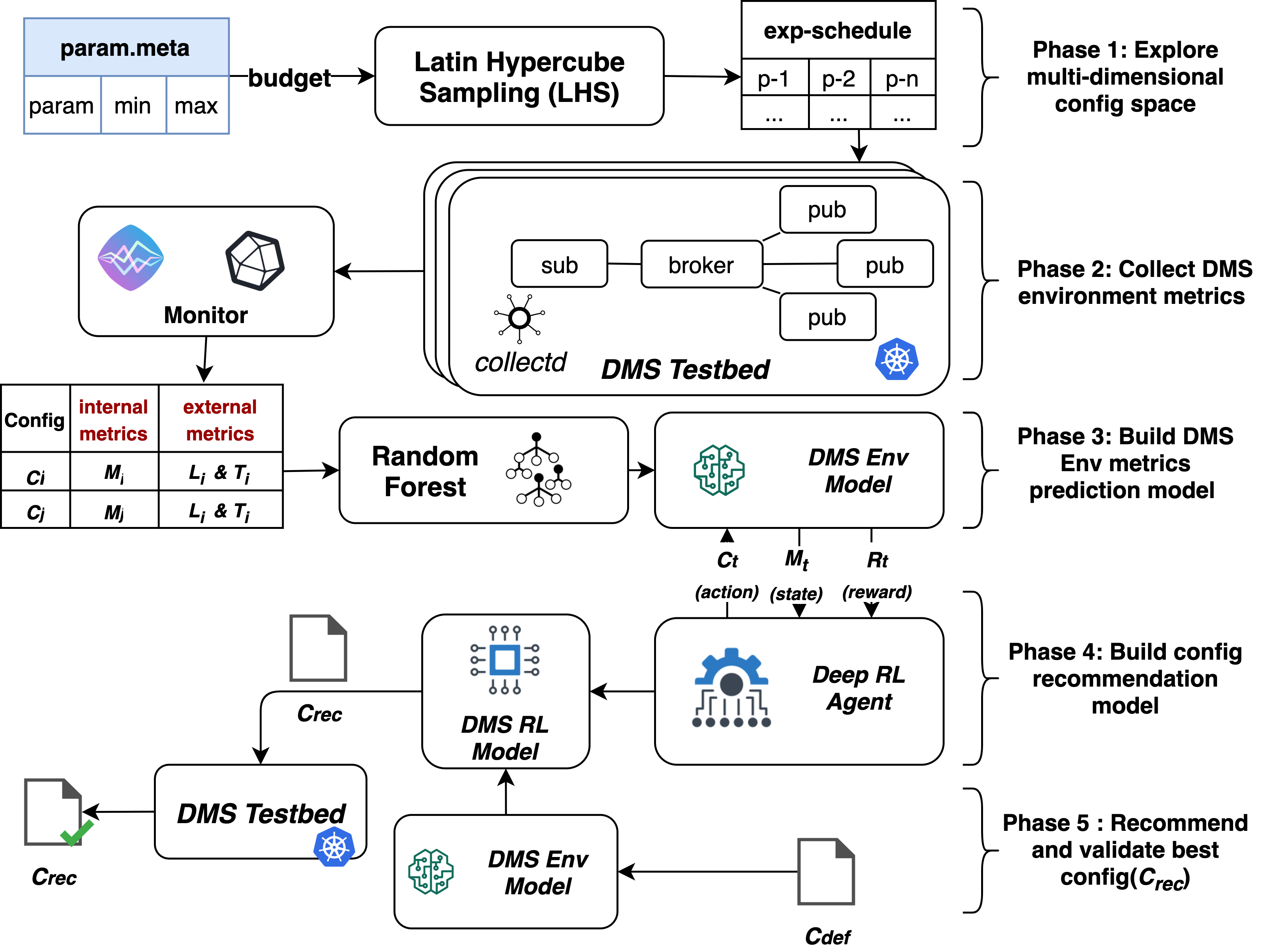}
    \vspace{-5pt}
    \caption{\textmd{DMSConfig Workflow and Architecture}}
    \label{fig:arch}
\end{figure}

We have set up the CCDMS problem formally. In this section, we introduce our approach DMSConfig, an automatic configuration optimization tool for distributed message systems. The architecture of DMSConfig is shown as figure~\ref{fig:arch}. In this research on DMSConfig, we make the following assumptions: (1) all producers (or brokers) share the same configuration; (2) application workload and topology remain static when serving a tuning request; (3) all connections in the environment have identical bandwidth; (4) the emulation testbed is equipped with the same hardware as that of the production environment. Next, we introduce the workflow and architecture of DMSConfig.



\subsection{Identifying Key Parameters} As mentioned in the first section, the fundamental challenge of the CCDMS problem sprang from its high-dimensional configuration space. Although there are considerable adjustable parameters in Kafka, only a small number of them have noticeable influence on application performance. Therefore, the purpose of this stage is to filter out unimportant parameters, which can help avoid the curse of dimensionality and improve model accuracy in latter steps. We identify vital parameters by the following steps: (1) refer Kafka manual and preceding benchmark results\cite{le2017performance, wang2015kafka} to roughly select some candidates; (2) evaluate the selected parameters using controlled experiments with the desired workload, and then observe performance variation to further prune redundant parameters. In the end, we identified the following list of important Kafka parameters and more details are available in Table~\ref{tab:param}:
\vspace{-5pt}
\begin{table}[H]
\centering
\def\arraystretch{1.1}%
\begin{tabular}[width=\textwidth]{|l|l|l|l|}
\hline
Parameter                   & Default & Range in DMSConfig   \\ \hline
num.network.threads         & 3       & 1-20                 \\ \hline
num.io.threads              & 8       & 1-24                 \\ \hline
queued.max.requests         & 500     & 50-5000              \\ \hline
socket.receive.buffer.bytes & 100KB   & 10-200KB             \\ \hline
socket.send.buffer.bytes    & 100KB   & 10-200KB             \\ \hline
socket.request.max.bytes    & 100MB   & 10-300MB             \\ \hline
buffer.memory               & 32MB    & 2-96MB               \\ \hline
batch.size                  & 16KB    & 4-256KB              \\ \hline
linger.ms                   & 0       & 0-100ms              \\ \hline
compression.type            & none    & none/snappy/gzip/lz4 \\ \hline
\end{tabular}%
\caption{\textmd{Identified Key Parameters}}
\label{tab:param}
\end{table}
\vspace{-0.5cm}

\begin{enumerate}[topsep=1pt]
\item \emph{num.network.threads:} controls the maximum number of threads in the broker used to listen or respond to network requests.
\item \emph{num.io.threads:} determines the number of I/O threads in the broker to process network requests.
\item \emph{queued.max.requests:} protects the broker from being overwhelmed by high-speed producer data-flows by limiting the maximum number of requests per broker. Network threads are blocked once this threshold is triggered.
\item \emph{socket.receive.buffer.bytes:} indicates the read buffer size of a socket connection, which controls the scale of ingress traffic.
\item \emph{socket.send.buffer.bytes:} determines the socket write buffer size. A large buffer size helps broker to tolerate bursty egress traffic.
\item \emph{socket.request.max.bytes:} sets the upper-bound of a single socket request to prevent the broker from being monopolized by large requests.
\item \emph{buffer.memory:} restricts the maximum allocatable memory for the producer for buffering. If records are sent faster than they can be transmitted to the server, then this buffer space will be consumed.
\item \emph{batch.size:} specifies the size of a single message batch. Micro-batch is an effective mechanism for improving application throughput where messages are coalesced and sent in small chunks. This parameter enables batch processing to improve latency as it prevents applications from frequent system calls and network stack traversal. The overhead is that increasing the waiting time of messages in memory may degrade latency. 
\item \emph{linger.ms:} designates a delay before a batched request is transmitted. In Kafka, messages are dispatched as long as one of the conditions in \emph{linger.ms} and \emph{batch.size} is met.
\item \emph{compression.type:} This parameter denotes the compaction method used by the producer, which includes four options: \emph{uncompressed}, \emph{gzip}, \emph{snappy}, and \emph{lz4}.
\end{enumerate}


\subsection{Testing Samples Generation} To expose the relationship between parameter combinations and performance as much as possible within a limited operation cost, we need an efficient sampling method to guarantee that the training set covers the configuration space uniformly. The Latin Hypercube Sampling(LHS)\cite{iman2014atin}, as a representative space-filling sampling technique, satisfies our demand in this sense. Specifically, LHS divides the range of each parameter into $N$ equally probable intervals, where $N$ denotes the number of desired samples. Then, for each parameter, it performs a stochastic sampling within a randomly selected interval. By iteratively performing random sampling without replacement, LHS ensures that values within the same interval at most appear once, thus evenly covering the sampling space.



\subsection{Benchmark and Data Collection} 
\label{subsec:benchmark}
Data collection is one of the most critical procedures for data-driven approaches as it is highly-coupled to model accuracy. However, a practical challenge we confront is how to collect data efficiently given the limitations on operational budget and accessible hardware resources. Compared to tasks in monolithic applications, this is harder for distributed systems like DMS because (1) they usually occupy more physical resources; (2) collected data may be noisy due to unpredictable network jitter; (3) the startup and cool-down time is relatively longer. Although LHS can represent the configuration space using much fewer samples than na\"ive grid or random sampling methods, evaluating 1000 configurations, for instance, still takes 25 hours, assuming we run 90 seconds to observe reliable performance metrics for each trial. If this is multiplied by the number of combinations of workload patterns and application topologies, the data collection phase would take a few weeks, which is certainly unacceptable. However, we argue that this cost can be linearly reduced using a container-based emulation testbed. Specifically, we use container virtualization technology to isolate the resources of physical nodes to parallelize the experimental process. Below we illustrate the feasibility of parallelizing experiments.

\begin{figure}[!htbp]
    \centering
    \includegraphics[width=0.85\columnwidth]{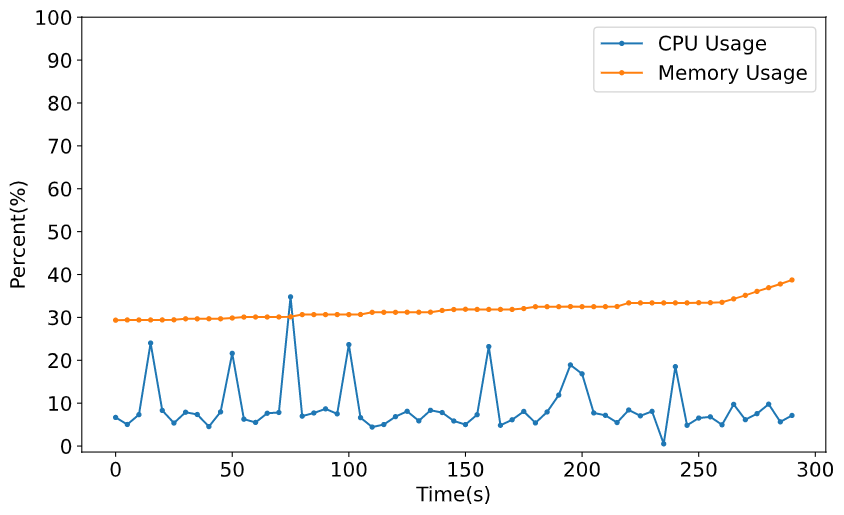}
    \caption{\textmd{CPU and Memory utilization rate of a Kafka server(8CPUs, 16GB RAM); the experiment is conducted with 10 producers(4CPUs, 8GB RAM), and 1 consumer(2CPUs, 4GB RAM); messages are published and subscribed at unlimited speed.}}
    \label{fig:res_utl}
\end{figure}

As claimed in \cite{TenRules35:online}, the Kafka server is not compute-intensive, so in most cases, the CPU and memory of physical nodes are not fully utilized in such benchmark tasks. To justify this and make reasonable resource allocation decisions for Kafka server, we performed a pressure test for a Kafka server in which it continuously processes high-frequency requests from 10 producers and consumers. As shown in Figure \ref{fig:res_utl}, the CPU and memory utilization rate of Kafka server is lower than 10\%, and 40\% percent, respectively, which inspires us to containerize benchmark applications. Docker container is a lightweight virtualization solution that uses Linux namespace and CGroup to isolate container processes and physical resources (CPU, memory, I/O devices) so that each application can run independently without interference by noisy neighbors. Also, bandwidth contention between applications augments data noise, consequently hindering testing multiple configurations simultaneously. However, these issues can be overcome using a container-based environment where the container engine creates a virtualized network interface for each container; thereby, we can employ some traffic shapers (e.g., Linux TC) to emulate span-node communication locally. As we execute containerized benchmark applications in a distributed manner, another practical problem we encountered is that deploying and managing large-scale short-live containers manually is tedious and laborious. Kubernetes(k8s) comes into our sight under this circumstance, as it is a widely accepted container orchestration platform for distributed applications. 


Specifically, for each test, DMSConfig first generates configuration files for participating entities on the k8s master node, then copies them to a labeled remote worker using the "scp" command. On worker nodes, containers mount required configuration files from the host for future benchmark tasks. We leverage Kubernetes Python API to define container meta information and interact with the Kubernetes cluster. Once the containers are deployed, DMSConfig asynchronously executes the startup commands on each and redirects container I/O to the master node. By then DMSConfig can monitor experiment progress and record logs centrally. Note that each experiment is orchestrated by an independent thread on the master. When the specified experiment time is reached, the thread immediately terminates the containers that it supervises.

We capture three categories of data including Kafka server states, container statistics, and producer performance (throughput and latency). The first two types are collectively termed as Kafka internal metrics, and the third is external metrics. Internal metrics reflect Kafka runtime behavior under specified input properties (workload, application topology, and resource distribution), used as states of our intelligent configurator in the configuration tuning phase to infer parameter adjustment decisions. External metrics directly characterize application performance. 
We leveraged Collectd \cite{forster2012collectd} to collect the system metrics as well as the docker container metrics. Collectd is a popular tool for gathering system and application statistics, and it offers pluggable interfaces for subsequent data storage and visualization.  We implemented a custom Collectd plugin called  \textit{Kafka-Container-Collectd} to monitor Kafka metrics running inside the docker containers. In DMSConfig, Collectd daemons are distributed over all physical hardware nodes in our cluster.  The \textit{Kafka-Container-Collectd} plugin periodically probes Kafka server containers on the host and queries the Kafka JMX agent for runtime metrics. Similarly, we used Collectd plugin \textit{Container-Collectd} which accesses container states through the Docker Python API. Query results are then relayed to an InfluxDB database running on the master node through a Rabbitmq messaging bus. As for external metrics, we redirected outputs of Kafka producer performance test application\cite{kafkagit} to log files on the master node and parse them after tests. We uniformly set the sampling frequency to 5 seconds, and select the 90th percentile of each metric as the effective measurement. For cumulative metrics, we calculate and store the difference between two consecutive measurements. 

In summary, we empirically select 7 internal metrics as shown below to describe Kafka container status.
\begin{enumerate}[topsep=2pt]
\item\emph{container.blkio.io\_service\_bytes}: Indicates the number of bytes read and written by the cgroup.
\item\emph{ container.cpu.usage\_usermode}: The amount of time the CPU was used executing tasks in user space.
\item\emph{container.cpu.usage\_kernelmode}: The amount of time the CPU was busy in kernel space.
\item\emph{ container.memory.usage\_total}: The total memory the container is using.
\item\emph{ kafka.produce.request.per\_sec}: The number of producer requests per second.
\item\emph{ kafka.produce.request.total\_time}: Total time (in ms) to serve the specified producer request.
\item\emph{ kafka.produce.request.temporary\_bytes}: Temporary memory used for message format conversions and decompression.  
\end{enumerate}

\subsection{DMS Simulator Training}
To build the DMS simulator that predicts the above Kafka internal state and performance metrics according to input configurations, we employ the random forest regression (RFR) algorithm to train the performance prediction model. Random forest is a non-parametric statistical learning technique that utilizes an effective ensemble-learning-based method that essentially performs a bootstrap aggregation (bagging) in which the final assessment decision is determined by the mean regression of a multitude of decision trees~\cite{liaw2002classification}. A decision tree is built by splits of nodes to adopt heuristic methods to partition the feature space. When splitting a node during the construction of a random forest, the forest takes a random subset of all features in the current state and chooses an effective slicing point. As each tree of the forest is constructed based on the bootstrap resampling strategy on the original training set and input features, RFR performs well with a small data set and even high-dimensional feature space. 


For given workload $W_i$, topology $T_i$, and resource profile $R_i$, we train an independent RFR model whose input is a testing configuration $C_t$, and the output is predicted Kafka metrics $S_t$. Formally, the regression model acts as an implicit function $f$ as defined in the equation ~\ref{eq:regression_model}:

\begin{equation}
\label{eq:regression_model}
\begin{aligned}
& S_t \leftarrow f_{\{W_i, T_i, R_i\}}(C_t)
\end{aligned}
\end{equation}

\begin{figure}[H]
    \centering
    \includegraphics[width=0.85\columnwidth]{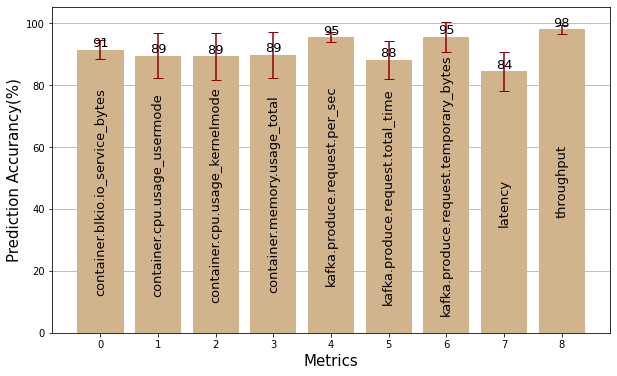}
    \caption{\textmd{Prediction accuracy of Kafka metrics among 9 use cases (see section~\ref{sec:exp}.})}
    \label{fig:acc}
\end{figure}

Empirical results reveal that RFR achieves 84\%-98\% prediction accuracy, as shown in Figure~\ref{fig:acc}, regarding selected Kafka metrics over 9 use cases(see section~\ref{sec:exp}), where 1,000 training samples are used for each scenario. Hence, we demonstrate that the obtained regression models are reliable enough for the DMS simulator.

\subsection{Configuration Tuning using DDPG}
When solving the parameter optimization problem, most traditional heuristic approaches (such as genetic algorithm, random search, evolutionary algorithm, etc.) make a fundamental assumption that the parameter space is discrete, but this is not always true for DMS. To optimize tuning in continuous space, we train an intelligent configuration tuner leveraging the DDPG algorithm, which takes the continuity of the search space into account and enhances configuration improvement direction deterministically according to the current state. DDPG is an policy-policy deep reinforcement learning algorithm, combining Deep Q learning networks (DQN)\cite{mnih2013playing} and Actor-Critic \cite{konda2000actor} methods. Specifically, it inherits from DQN and includes: (1) adopting a deep neural network to approximate the Q function, (2) using the Experience Replay mechanism to break the empirical data correlation caused by MDP. Actor-critic methods are Temporal Difference (TD) methods, where the TD approximates the current estimate based on the previously learned estimate. Thus, Actor-critic methods have a separate memory structure to explicitly represent the policy independent of the value function. Actor takes suggestion from Critic and generates a tuning action, environment that deploys the tuning action, and a reward value based on the performance change on the updated configuration. Critic criticizes the actions made by the actor and estimates the state value function. Thanks to the Actor-Critic method, DDPG can learn in high-dimensional continuous action spaces because it selects actions based on the policy gradient method instead of examining Q-value for every action as DQN does. Next, we describe more implementation details of DDPG and our design of the reward function.

\subsubsection{\textbf{DDPG in DMSConfig}}
DDPG constructs four neural networks in light of the structural features of the Actor-Critic method with the Temporal Difference (TD) estimate principle, where both Actor and Critic have an online ($\theta^u, \theta^Q$) network and a target($\theta^{\mu'}, \theta^{Q'}$) network respectively. Each pair of online and target network share identical architectures. DDPG adopts a "soft update" approach to slowly blend the weights of online networks to corresponding target networks weights. The policy function $a_i = \mu(s_i|\theta^\mu)$ is substituted by the actor-network $\theta^{\mu}$, which estimates the action $a_i$ that agent should perform in state $s_i$. To avert falling into the local optimum during training, we add Gaussian noise to action $a_i$ so that agent has opportunities to explore unknown searching areas. While interacting with the environment, the agent records the state transition $T_i=<s_ i, a_ i, r, s_ {i + 1}>$ into a fixed size Experience Replay Buffer ($ERB$). When the Actor and Critic networks need to be updated, DDPG randomly extracts a minibatch $N=\{T_1, T_2, ..., T_n\}$ from $ERB$. 
Then for every record $T_i$, the online critic-network $\theta^Q$ takes $s_i$ and $a_i$ as inputs, and outputs a Q value, $Q(s_i, a_i|\theta^Q)$, indicating the benefit of executing $a_i$ in state $s_i$. Therefore, the actor policy should be updated along the direction instructed by the critic-network to improve Q-values and expose better actions, which is formally representing as equation ~\ref{eq:gradient}:
\begin{equation}
\label{eq:gradient}
\begin{aligned}
& \nabla_{\theta^\mu} J \approx \frac{1}{N} \sum_{i} \nabla_a Q(s, a|\theta^Q) |_{s=s_i, a=\mu(s_i)} \nabla_{\theta^\mu}\mu(s|\theta^\mu)|_{s=s_i}
\end{aligned}
\end{equation}
, where $J$ denotes all possible policies. As for the critic-network, it conducts back-prorogation to update weights of the neural network by minimizing the TD-error. Equation ~\ref{eq:td-error} shows the TD-error as mean square error between target Q-value and estimated Q-value:

\begin{equation}
\label{eq:td-error}
\begin{aligned}
& L(\theta^Q) = \frac{1}{N} \sum_{i} (y_i - Q(s_i, a_i|\theta^Q)^2)
\end{aligned}
\end{equation}

, where $y_i=r_i + \gamma Q'(s_{i+1}, \mu'(s_{i+1}|\theta^{\mu'})|\theta^{Q'})$, denotes the target-Q-value.
\vspace{-5pt}

\begin{figure}[H]
    \label{fig:actor-critic}
    \centering
    \includegraphics[width=\columnwidth]{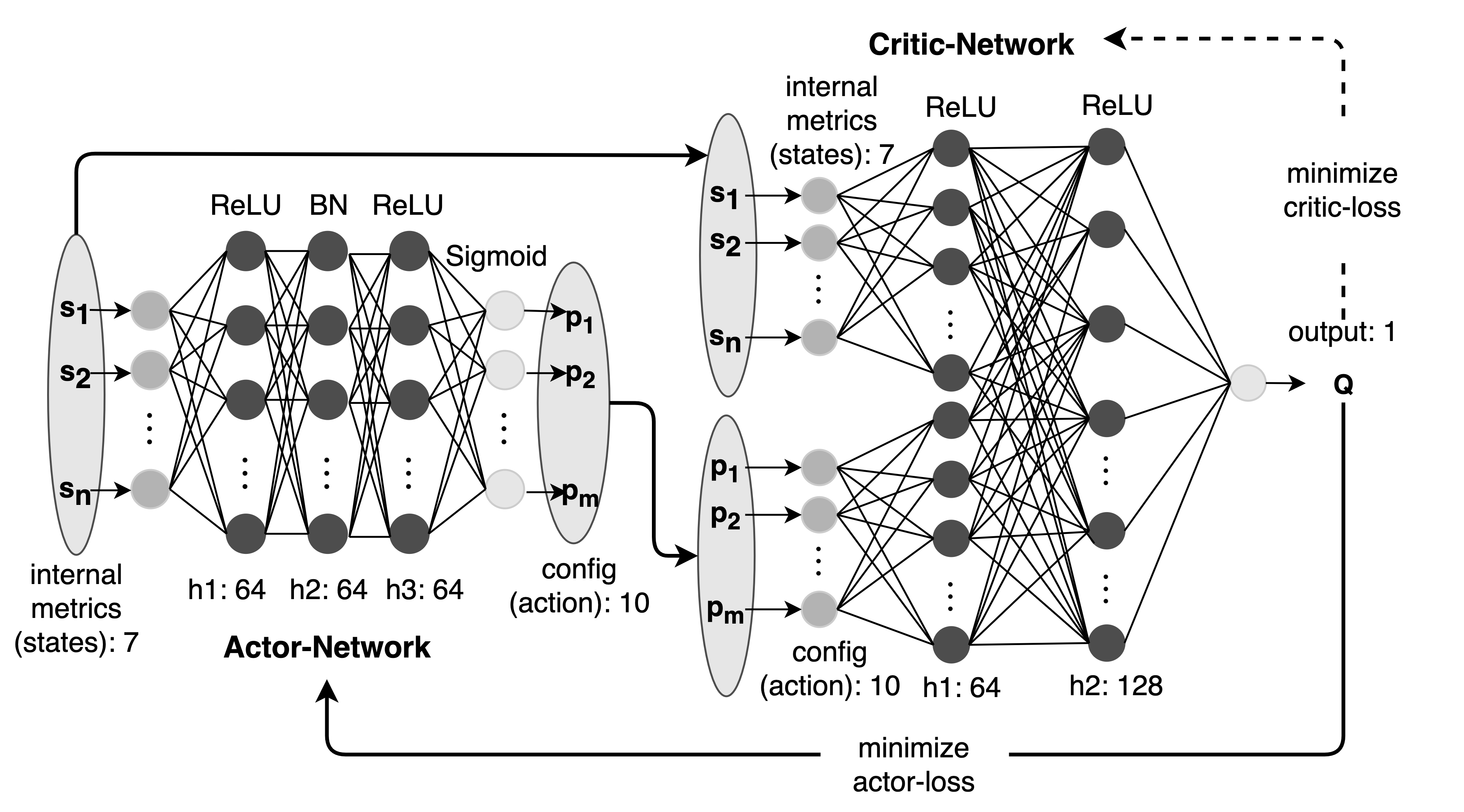}
            \vspace{-10pt}

    \caption{\textmd{Architecture of DDPG}}
    \label{fig:ddpg}
\end{figure}
The actor-critic networks of DDPG is defined in Figure ~\ref{fig:ddpg}. Specifically, the actor-network contains two hidden layers with the ReLU activation function, a batch normalization(BN) layer, and an output layer activated by the Sigmoid function. As for the critical-network, state and action vectors are first accepted by two parallel ReLU layers, and then being concatenated with another ReLU layer. Moreover, we replace the $ERB$ in general DDPG with a prioritized $ERB$~\cite{schaul2015prioritized}, which is proven effective in accelerating the convergence of actor policy\cite{zhang2019end}. In short, the prioritized $ERB$ ranks transition samples from the minibatch based on the magnitude of TD-error, which significantly improves the diversity of samples. 


\vspace{-5pt}
\subsubsection{\textbf{Reward Function}}
Eq.~\ref{eq:td-error} and ~\ref{eq:gradient} suggest that the reward function is essential for DDPG navigating actor to update policy toward to the correct direction. Recalling Eq.~\ref{eq:obj}, our optimization objective is to maximize throughput under latency constraint; thus, the reward function has three missions: (1) navigate agents to chose actions that have higher throughput than that at the initial state; (2) reward agents when the throughput gets improved compared to the previous step; (3) prevents agents from exceeding latency boundary. 

The rationality of the reward function mission setup has been proofed in ~\cite{zhang2019end,li2019qtune} that when the reward function simultaneously compares current throughput with both of the initial state $T_0$ and the previous step $T_{t-1}$, the actor policy gains fastest convergence and best performance . Therefore, we inherit the same principle in ~\cite{zhang2019end,li2019qtune} and extend it to scenarios with strict latency constraints. Formally, let $T_t$ and $L_t$ be throughput and latency responded by environment at timestamp $t$, respectively. Since configurations (actions), internal metrics (states) and external metrics(throughput, latency) have been normalized when we train the DMS simulator, we directly compute performance changes as follows:
\vspace{-10pt}
\begin{equation}\notag
\begin{aligned}
    \left \{
    \begin{array}{lc}
    \Delta T_{t,0} = T_t - T_0\\
    \Delta T_{t,t-1} = T_t - T_{t-1} \\
    \Delta L_{t,c} = L_t - L_c
    \end{array}
    \right.
\end{aligned}
\end{equation}
, where $L_c$ indicates latency constraint constant. The overall reward function is designed as:
\vspace{-10pt}
\begin{equation}
\label{eq:reward}
\begin{aligned}
r = \left \{
\begin{array}{lc}
(-1-\Delta L_{t,c})^{\lfloor \frac{L_t}{L_c} \rfloor} ((1+\Delta T_{t,0})^2 -1) |1+\Delta T_{t,t-1}|, \Delta T_{t,0} > 0 & \\
(1+\Delta L_{t,c})^{\lfloor \frac{L_t}{L_c} \rfloor} ((1-\Delta T_{t,0})^2 -1) |1-\Delta T_{t,t-1}|, \Delta T_{t,0} \leq 0
\end{array}
\right .
\end{aligned}
\end{equation}
In summary, when throughput is better than that at the initial state(throughput of the default configuration) and latency satisfies the constraint, the latency penalty term is omitted; otherwise, the throughput part is shadowed by a negative latency multiplier. The more latency oversteps the boundary, the stronger the penalty. Similarly, if the throughput is worse than that in the initial state, and the latency boundary is crossed, we augment the overall penalty.

%% file: experiment.tex
\vspace{-5pt}

\section{Evaluation}
\label{sec:exp}


In this section, we evaluate DMSConfig under 9 Kafka use cases and compare its performance with three mature configuration optimization tools and the default one provided by Kafka vendor. Besides, we examine 7 different levels of latency constraints for each use case to answer whether DMSConfig can successfully balance latency and throughput and obtain maximum profits meanwhile.

\begin{table}[H]
\centering
\def\arraystretch{1.1}%
\begin{tabular}{|c|c|c|c|c|}
\hline
No. &
  \begin{tabular}[c]{@{}c@{}}Producer \\ Number\end{tabular} &
  \begin{tabular}[c]{@{}c@{}}CPU \\ Number\end{tabular} &
  \begin{tabular}[c]{@{}c@{}}Bandwidth \\ (Gbps)\end{tabular} &
  \begin{tabular}[c]{@{}c@{}}Message Size \\ (KB)\end{tabular} \\ \hline
1 & 1  & 2 & 1.0   & 0.1 \\ \hline
2 & 1  & 2 & 1.0   & 1.0 \\ \hline
3 & 1  & 2 & 1.0   & 4.0 \\ \hline
4 & 1  & 1 & 1.0   & 1.0 \\ \hline
5 & 1  & 4 & 1.0   & 1.0 \\ \hline
6 & 1  & 2 & 0.1   & 1.0 \\ \hline
7 & 1  & 2 & 0.5   & 1.0 \\ \hline
8 & 3  & 2 & 1.0   & 1.0 \\ \hline
9 & 10 & 2 & 1.0   & 1.0 \\ \hline
\end{tabular}
\caption{\textmd{Use cases, where test 2 is regarded as the referential test.}}
\label{tab:schedule}
\end{table}
\vspace{-1cm}
\textbf{Experiment Environment} Our experiment environment comprises 10 AMD Opteron(tm) Processor 4170 HE nodes with 2.10GHz CPU speed, each with 12 physical cores, and 32GB memory; and their software specifications are as follows: Ubuntu 16.04 OS, Linux 4.4.0-157-generic kernel, JDK V1.8.0, Kubernetes V1.18, and Docker V19.03. All of the bare-metal machines are connected via a high-speed 1Gbps LAN. Moreover, as depicted in Section~\ref{sec:dmsconfig}, we perform benchmarking tasks on a Kubernetes cluster, which contains a master node and nine worker nodes that communicate with each other through Flannel CNI. 

\textbf{Benchmark Application} We leverage the official Kafka performance evaluation application\cite{kafkagit} to create streaming workloads and capture corresponding throughput and latency. Unless otherwise specified, each containerized Kafka server has 4CPUs and 8GB RAM. The producer is equipped with 2CPUs and 4GB RAM. As for the subscriber, we deploy 1CPU and 4GB RAM. It should be noted that since this paper primarily focus on optimizing throughput for latency-aware producers, we deploy a single broker and one subscriber for all tests. Producers and consumers send and consume messages at unrestricted frequencies, respectively. Additionally, we consider reliable communication only so that the Kafka server is required to acknowledge every message dispatched by producers.

\textbf{Baseline Techniques} Sequential model-based Bayesian optimization is the mainstream scheme in the field of black-box optimization whose internal implementation is mainly based on (1) a surrogate model for predicting validation loss (2) and a heuristic acquisition function for selecting testing points sequentially.  In contrast, we leverage the reward mechanism and policy gradient method of DDPG, rather than a prior-knowledge-base acquisition function, to lead a tuner incrementally improve the quality of selected parameters. Hyperopt\cite{bergstra2013hyperopt}, Optuna\cite{akiba2019optuna}, and SMAC\cite{hutter2011sequential} are three representative implementations of SMBO. Hence, we use them as baselines and briefly summarize their basic information and working principle as below:
\begin{itemize}
    \item Hyperopt is a powerful hyper-parameter tuning instrument based on Bayesian optimization. We realize this optimizer using its open source Python library\footnote{\url{http://hyperopt.github.io/}} with Tree Parzen Estimators as the fitting model and expected improvement as the acquisition function.
    \item SMAC is also a Bayesian optimization solver. Likewise, our implementation is based on the SMAC3 \footnote{\url{https://github.com/automl/SMAC3}} library with default algorithm settings(Random Forests as the surrogate loss prediction model, logarithm expected improvement as the acquisition function).
    \item Optuna: Optuna is regarded as the next-generation automatic hyper-parameter optimization software, which uses the same optimization method as that of Hyperopt but equips an additional pruner to discard unpromising trials to improve sampling quality when choosing testing points. We implement experimental codes using their Python API\footnote{\url{https://github.com/optuna/optuna}}.
\end{itemize}

To make fair comparisons, we set the reward function, as shown in Eq.~\ref{eq:reward}, as the objective function of all baseline algorithms. In addition, because it is extremely time-consuming to carry out online assessments on the practical testbed directly, we adopt the DMS simulator trained in section~\ref{sec:dmsconfig} to generate DMS external metrics needed for the objective function. 

\textbf{Experiment Design}
Based on the previous Kafka benchmark experience\cite{le2017performance,wang2015kafka}, we design 9 test cases to simulate various scenarios from the aspects of message size, producer computing capacity, network bandwidth, and producer quantity. We perform controlled experiments using configurations recommended by each tuner under these use cases, and run each for 90 seconds to obtain reliable performance metrics. To understand baseline performance and validate the effectiveness of DMSConfig, we execute the Kafka default configuration, as shown in table~\ref{tab:param}, with the same design of experiments. Table~\ref{tab:schedule} shows experiment settings where the second is considered as the referential test for studying influences of different variables. In addition, we draw latency boundaries based on the default configuration and compute the latency constraint factor $lcf$ as the follow:
\begin{equation}\notag
    lcf = \frac{L_{default}}{L_{c}}
\end{equation}
, where $L_{default}$ refers to the latency earned by the default configuration, and $L_c$ is the latency restriction set for tuners. Particularly, $lcf=0$ denotes there is no restriction for latency.

\begin{figure}[H]
    \centering
    \includegraphics[width=0.85\columnwidth]{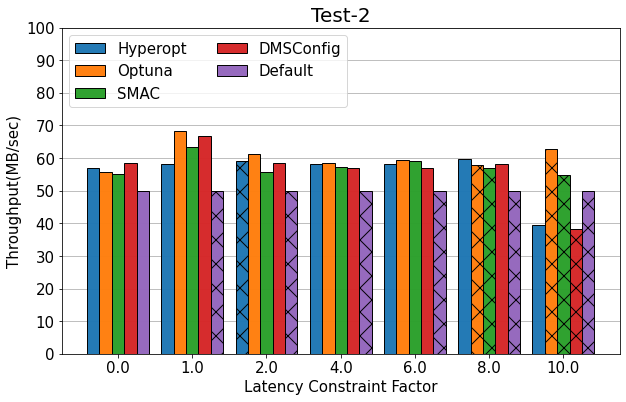}
    \vspace{-10pt}
    \caption{\textmd{Referential Test for controlled experiments.}}
    \label{fig:test2}
\end{figure}

\begin{figure}[ht]
\centering
\subfloat{%
\centering
\includegraphics[width=0.85\columnwidth]{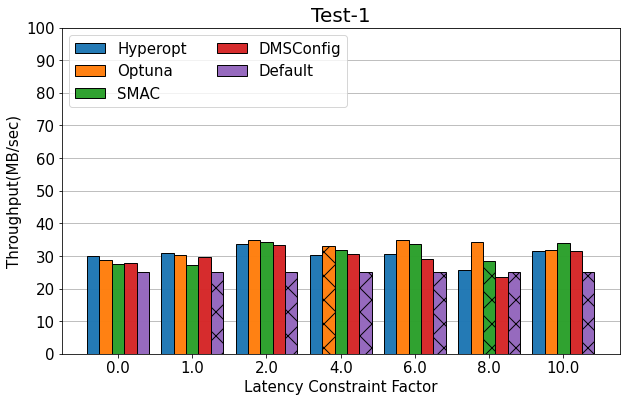}
}
\quad
\subfloat{%
  \centering
  \includegraphics[width=0.85\columnwidth]{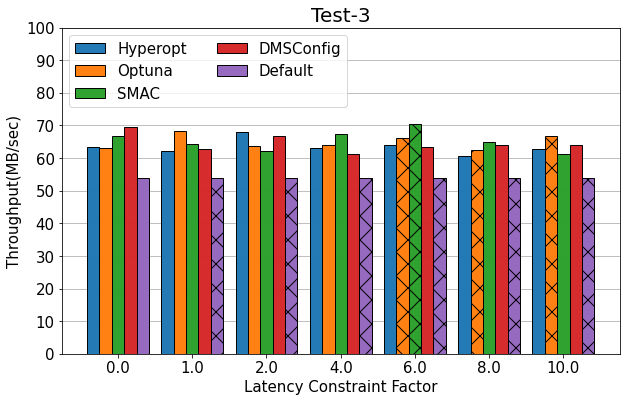}
}
        \vspace{-10pt}
\caption{\textmd{Performance comparison under different message lengths(Test-1:0.1KB, Test-2:1KB, Test-3:4KB).}}
\label{fig:message-test}
\end{figure}

\begin{figure}[ht]
\centering
\subfloat{%
  \centering
  \includegraphics[width=0.85\columnwidth]{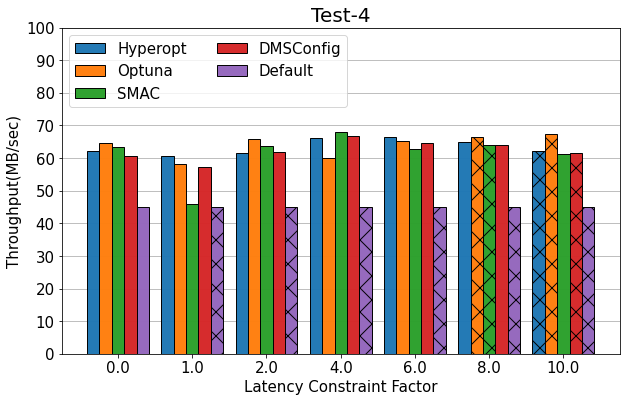}
}
\quad
\subfloat{%
  \centering
  \includegraphics[width=0.85\columnwidth]{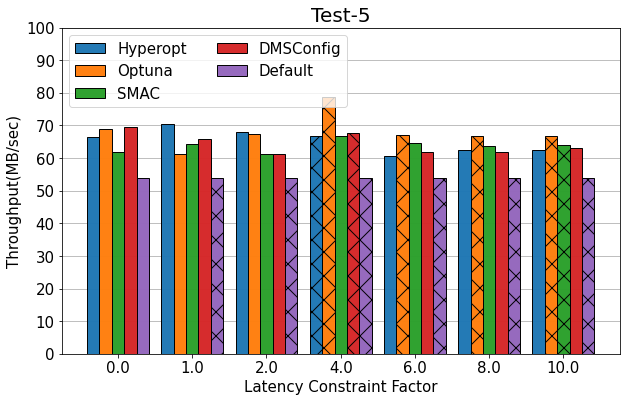}
}
        \vspace{-10pt}

\caption{\textmd{Performance comparison under different number of producer CPUs(Test-4:1, Test-2:2, Test-5:4).}}
\label{fig:cpu-test}
\end{figure}

\begin{figure}[ht]
\centering
\subfloat{%
  \centering
  \includegraphics[width=0.85\columnwidth]{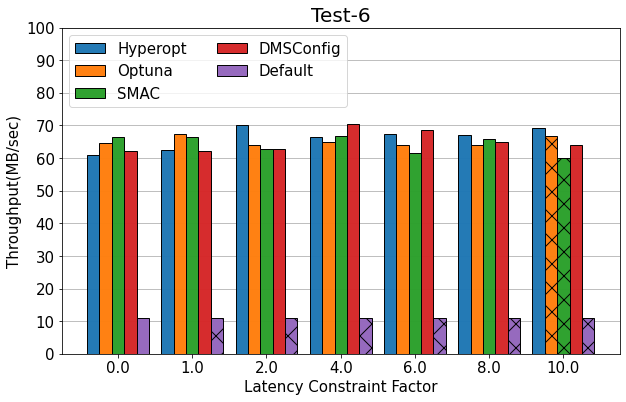}
}
\quad
\subfloat{%
  \centering
  \includegraphics[width=0.85\columnwidth]{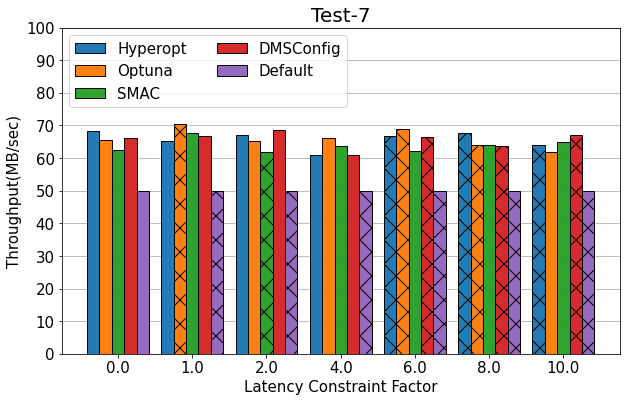}
}
        \vspace{-10pt}

\caption{\textmd{Performance comparison under various bandwidth settings(Test-7: 0.1Gbps, Test-8: 0.5Gbps, Test2: 1Gbps)}}
\label{fig:bw-test}
\end{figure}

\begin{figure}[ht]
\centering
\subfloat{%
  \centering
  \includegraphics[width=0.85\columnwidth]{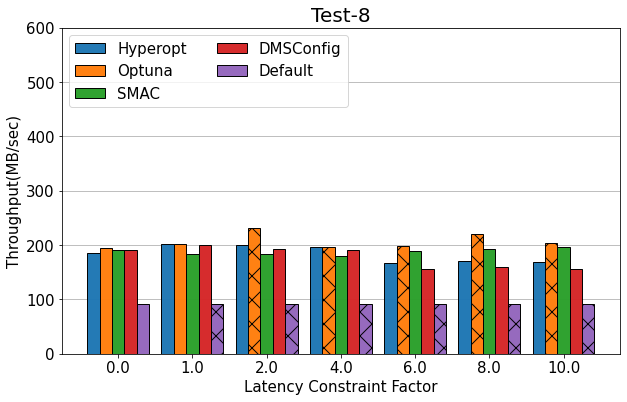}
}
\quad
\subfloat{%
  \centering
  \includegraphics[width=0.85\columnwidth]{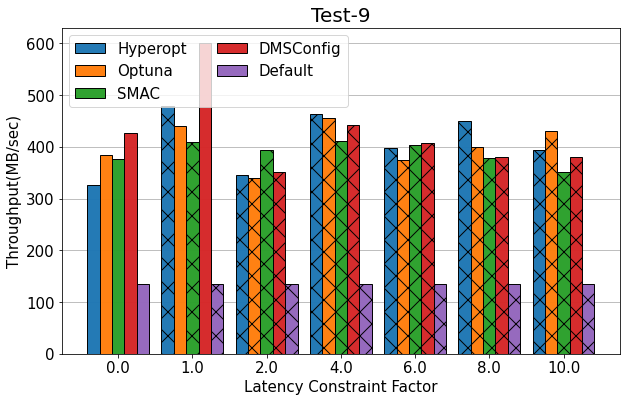}
}
        \vspace{-10pt}

\caption{\textmd{Performance comparison under different number of producers(Test-8:3, Test-2:1,Test-9:10).}}
\label{fig:pub-test}
\end{figure}

\textbf{Experiment Results} In this section, we demonstrate our experiments results in terms of two evaluation metrics: throughput and latency violations. It is worth mentioning that the throughput mentioned here reflects the number of message bytes delivered successfully by producers, but it does not necessarily match the actual bandwidth usage due to message compression and concurrent communication under some settings. In the multi-producer cases (Test 8, 9), we sum the throughput accomplished by each producer as the overall throughput of the system. In addition, in the bar chart shown in figure~\ref{fig:message-test}-\ref{fig:pub-test}, we texture the bar if a configuration fails to meet the specified latency constraint. In order to thoroughly understand the 
experiment results, we first analyze the performance of default configuration in different use cases, and then compare it with DMSConfig.
\begin{table}[H]
\centering
\begin{tabular}{|l|c|c|c|c|c|c|c|c|c|}
\hline
\diagbox[width=2cm]{$lcf$}{Imp(\%)}{Test} & 1 & 2 & 3 & 4 & 5 & 6 & 7 & 8 & 9 \\ \hline
0         & 8      & 16     & 27     & 33     & 27                         & 463    & 32                         & 106    & 218                         \\ \hline
1         & 16     & 32     & 14     & 26     & 20                         & 463    & 32                         & 117    & 347                         \\ \hline
2         & 32     & 16     & 22     & 35     & 12                         & 463    & 36                         & 108    & \cellcolor[HTML]{C0C0C0}161 \\ \hline
4         & 20     & 14     & 12     & 46     & \cellcolor[HTML]{C0C0C0}24 & 536    & 22                         & 106    & \cellcolor[HTML]{C0C0C0}229 \\ \hline
6         & 12     & 14     & 16     & 42     & 12                         & 518    & \cellcolor[HTML]{C0C0C0}32 & 69     & \cellcolor[HTML]{C0C0C0}203 \\ \hline
8         & -8     & 16     & 16     & 40     & 12                         & 481    & \cellcolor[HTML]{C0C0C0}26 & 72     & \cellcolor[HTML]{C0C0C0}184 \\ \hline
10 & 24 & \cellcolor[HTML]{C0C0C0}-24 & 18 & \cellcolor[HTML]{C0C0C0}35 & 16 & 481 & \cellcolor[HTML]{C0C0C0}34 & 68 & \cellcolor[HTML]{C0C0C0}183 \\ \hline
\end{tabular}%
\caption{\textmd{Throughput Improvement(DMSConfig VS. Default), gray cells indicate latency violation cases.}}
\label{tab:imp}
\end{table}
\vspace{-1cm}
\emph{Understand The Default Configuration} From the message size tests, see figure~\ref{fig:test2},~\ref{fig:message-test}, we can see that under the single producer setting, when the message size increases from 0.1KB to 1KB, the throughput is doubled (25MB/sec-50MB/sec); however, as the message continues growing to 4KB, the throughput only increases by 4MB/sec (50MB/sec-54MB/sec). This shows that under the default configuration, the 1KB high-speed data stream can almost make Kafka server fully loaded (throughput is about 50MB/sec), which is consistent with the performance evaluation results from Kafka community~\cite{ApacheKa30:online}. As for producer CPU tests, figure~\ref{fig:cpu-test} suggests the number of CPU may become performance bottleneck, because the producer with single core cannot use multithreading to process I/O requests. Besides, as can be noticed in figure~\ref{fig:bw-test}, in the low bandwidth environment(0.1Gbps), the performance of the default configuration is significantly cut off. However, in the case of 0.5Gbps, the throughput is equal to that of the referential test~\ref{fig:test2}, which indicates a single-producer will not cause network overload under current settings. However, when we run 10 producers simultaneously, as suggested in figure~\ref{fig:pub-test}, the overall throughput is approximately equal to the physical bandwidth. (Message compression is disabled in the default configuration, so this statement holds.)

\emph{Comparison between DMSConfig and Default} By comparing the configuration recommended by DMSConfig with the default one, we mainly answer two questions: 1) whether DMSConfig can suggest a better configuration even with latency constraints; 2) whether DMSConfig can make adaptive decisions to ensure optimization objectives when performance bottlenecks arise. We demonstrate comparison results in Table~\ref{tab:imp}, where shaded cells represent the recommended configuration fails to meet the latency constraint. As can be seen from Table~\ref{tab:imp}, DMSConfig achieved 8\%-463\% and 14\%-463\% higher throughput than the default configuration, respectively, over 9 experiments in loose latency constraint conditions ($lcf=0$, $lcf=1$). However, as we augment the restriction, latency violation occurs more frequently since the default configuration is designed for optimal latency, as depicted in \cite{TailLate0:online}, which is most evident in the case of $lcf = 10$. Overall, DMSConfig achieves 12\%-538\% throughput improvement over 7 levels of latency restrictions in 79.63\% of tests(total 63). Concerning the second question, we focus on Test 4 and 6 as they are situations where the default configuration suffers. By comparing Figures~\ref{fig:cpu-test} and ~\ref{fig:bw-test}, we can observe that DMSConfig can consistently sustain throughput at 58MB/sec-70MB/sec in CPU or bandwidth-constrained environment, which is almost the same as that of the control group(see Fig.~\ref{fig:test2}). In summary, we claim DMSConfig is an effective approach to locate promising configurations in latency-aware DMS.
\begin{figure}[H]
    \centering
    \includegraphics[width=0.85\columnwidth,height=4cm]{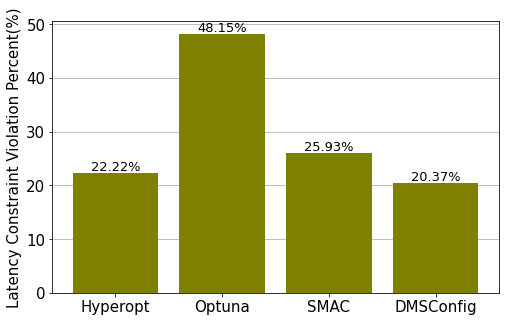}
        \vspace{-10pt}
    \caption{\textmd{Comparison of latency violation percent.}}
    \label{fig:lat_vio}
\end{figure}
\vspace{-20pt}
\vspace{-10pt}
\emph{Comparison between DMSConfig and Baseline Algorithms} In general, the difference in throughput between DMSConfig and the other three Bayasien-based baselines is trivial. Optuna is the most aggressive one in tracing a configuration with higher throughput; however, as can be noticed in Figure~\ref{fig:lat_vio}, it is also more likely to trigger latency constraint. In contrast, DMSConfig is more reliable in fulfilling latency restrictions with 20.37\% violation percent, which is lower than other approaches. 

%% file: related.tex
\section{Related work}
\label{sec:related}

The fundamental problem in automating the Pub/Sub system configuration is to tackle the high dimensional configuration parameter space and complex combinatorial relationships among configuration knobs, which is similar to other network systems such as database management systems, distributed data analytic platforms, and web servers. To that end we compare prior efforts in configuration management for different systems.

\subsection{Configuration Tuning optimization in network systems}
\label{subsection:Tuning_related}

Search-based methods solve the network systems configuration problem in a blackbox manner. Herodotou et al. integrated a recursive random search to tune the Hadoop cluster size ~\cite{Herodotou2011no}. Zhu et al. presented BestConfig to improve the  system throughput via a divide-and-diverge sampling and a recursive bound-and-search algorithm~\cite{zhu2017bestconfig}. Dou et al. employed a modified random embedding Bayesian search based algorithm to solve the configuration optimization problem on a search engine~\cite{ dou2020hdconfigor}. The search-based method's rudimentary problem is how to deal with a local search range and sample randomness at each search iteration. 
            

Prediction based methods establish performance prediction models after collecting different configuration sets. One of the challenges is that only a limited set of samples can be acquired under a given time constraint, especially for real production toolkits. Sarkar et al. enhanced projective sampling by incorporating two heuristics for performance prediction to sampling system cost efficiency ~\cite{sarkar2015cost}. Sullivan et al. employed an influence diagram formalism based method to predict the Berkeley DB performance under a given workload ~\cite{sullivan2004using}. There are a few studies using machine learning to build a prediction model and tune distributed data analytic systems such as random forest ~\cite{bei2015rfhoc}, support vector machine ~\cite{lama2012aroma}, K-means ~\cite{stewart2017dynamic}. 
            

Ranked based approaches reduce the training sets by ordering important parameters in discrete configuration space. Bao et al. proposed Autoconfig that first selects top $k$ essential elements from a prediction model, then utilizes a weighted Latin Hypercube Sampling and a search algorithm to optimize Kafka system throughput ~\cite{bao2018autoconfig}. Van et al. employed a ranked list to select the most impactful configuration parameters with a combination of supervised and unsupervised machine learning pipeline ~\cite{van2017automatic}. Nair et al. proposed a rank-based approach that ranks the difference between actual and predicted performance to optimize configurable systems~\cite{nair2017using}. Mahgoub et al. applied the ANOVA-based analysis to identify the key parameters that impact the throughput performance of several NoSQL databases ~\cite{mahgoub2017rafiki}. In short, obtaining the discrete space from a continuous space is complicated and needs external tuning. 
    

As prediction models are constructed, reinforcement learning (RL) is an efficient alternative way to find optimal configurations instead of traditional search algorithms. Vaquero et al. presented an automated approach to recommend the most appropriate configurations for Spark workloads ~\cite{vaquero2018auto}. Their methods include supervised machine learning-based metrics ranking mechanism and Q learning-based RL based on the previous workload. Zhang et al. proposed CDBTune that utilizes the DDPG based deep RL with a reward-feedback mechanism to do online-tuning on the cloud database system's continuous configuration space. CDBTune is trained with a limited number of samples to avoid collecting massive samples~\cite{zhang2019end}. Similarly, Li et al. extracted rich features of SQL queries, then the authors utilized a Double-State DDPG model that enables the actor-critic networks to tune the database configuration based on both the query vector and database states ~\cite{li2019qtune}.

Our work DMSConfig explores a near optimal configuration solution in continuous parameter space without splitting the space in discrete set in Pub/Sub stream processing software, which is the main difference from the above works. 
\subsection{Latency-aware constrained in Pub/Sub stream processing}

The DMSConfig aims to hold the latency constraint in the Kafka platform layer. None of the above works explicitly discussed latency constraint in the software platform layer; instead, we care about the latency constraint QoS awareness. The relationship between latency and throughput is inverse if the message queuing size is fixed. However, with more IoT devices being deployed with 5G, the scalability of legacy stream processing systems is challenged by meeting the tradeoff between throughput and latency. Some recent studies have investigated the pub/sub systems architecture component for latency-constrained distributed applications to reduce average response time and enhance outgoing message throughput. 

Gascon-Samson et al. proposed Dynamoth to optimize system-level and channel level load-balancing ~\cite{ gascon2015dynamoth}. The balancing enables the Redis Pub/Sub system to scale to arbitrary numbers of publishers, subscribers, and publications in the system level and cross channels in real-time to adapt to the workload. Multipub is a flexible latency-constrained Pub/Aub system ~\cite{gascon2017multipub}. It dynamically reconfigures the message delivery configuration within multiple-brokers to achieve a latency guarantee for the messages and cloud costs. Khare et al. proposed a framework that learns a latency prediction model on an edge broker and uses this model to balance the processing and data-dissemination load to provide the desired QoS for the latency-aware placement of topics on brokers ~\cite{khare2018scalable}. Balasubramanian et al. employed an additive-increase multiplicative-decrease based control algorithm that auto-tunes the Pub/Sub system batch size and the polling interval to optimize the input message load and solve to broker-side congestion that minimizes the latency ~\cite{balasubramanian2018auto}. Hasenburg et al. presented the latency and excess data dissemination tradeoff within Pub/Sub systems running in fog environments. The authors implemented a broadcast group that split the set of edge brokers into connected groups that use flooding for intra-group communication and a cloud relay for intergroup communication ~\cite{hasenburg2020managing}. 

Again, our work focuses on employing latency constraint to optimize Kafka software's system configuration, rather than topology configuration, job scheduling, or routing configuration. 


%% file: conclusion.tex
\section{Conclusion}
\label{sec:conclusion}
In this paper, we propose DMSConfig, an automated configuration tuning approach for latency-aware IoT message systems. DMSConfig casts the problem of constrained configuration optimization to an RL task and searches the near-optimal configuration leveraging the DDPG algorithm. We construct a regression model using RF to predict performance metrics and internal states of a DMS under specific software configurations. Our well-designed reward function leads the RL agent to learn to make wise configuration decisions in a high-dimensional continuous configuration space by iteratively interacting with the regression model. To overcome the burden of data collection cost, we develop a container-based full-stack emulation testbed to parallel experiments. Extensive experimental results reveal that the configurations identified by DMSConfig achieve 12\%-538\% throughput improvement, compared with the default one suggested by Kafka manual, under 7 different levels of latency restrictions. In addition, DMSConfig is able to guarantee application performance under resource-constrained environments by making wise configuration recommendations. We also verified that DMSConfig earns analogous throughput performance compared with three SMBO-based state-of-the-art parameter optimization frameworks, but delivers the most reliable latency guarantees. 

In future work, we plan to (1) employ Generative Adversarial Network(GAN) to alleviate the data collection cost further and improve the accuracy of our environment prediction model; (2) integrate the diversities of DMS workloads, topology, and resource allotments into the environment model to enable RL agent competent dynamic configuration recommendation tasks; (3) explore more RL reward functions for improving quality of suggested configurations.

%% file: acknowledgements.tex
\begin{acks}
This work is supported by a funding from Cisco.  Any opinions, findings, and conclusions or recommendations expressed in this material are of the author(s) and do not necessarily reflect the views of the sponsors.
\end{acks}